\documentclass[traditabstract]{aa}  
\usepackage{graphicx,amsmath,hyperref,color}
\usepackage{txfonts}

\begin{document}

   \title{A Pair of Early- and Late-Forming Galaxy Cluster Samples: \\ a Novel Way of Studying Halo Assembly Bias Assisted by a Constrained Simulation}
   \titlerunning{A Pair of Early- and Late-Forming Clusters}

   \author{Yen-Ting Lin\inst{1}\thanks{ytl@asiaa.sinica.edu.tw} \and Hironao Miyatake\inst{2,3} \and Hong Guo\inst{4} \and Yi-Kuan Chiang\inst{1} \and Kai-Feng Chen\inst{5} \and Ting-Wen Lan\inst{6,1}  \and \\ Yu-Yen Chang\inst{7}}

   \institute{Institute of Astronomy and Astrophysics, Academia Sinica  (ASIAA),  Taipei 10617, Taiwan \and
             Kobayashi-Maskawa Institute for the Origin of Particles and the Universe (KMI), Nagoya University, Nagoya, 464-8602, Japan \and
       Kavli Institute for the Physics and Mathematics of the Universe, The University of Tokyo Institutes for Advanced Study, The University of Tokyo, Chiba 277-8583, Japan \and
          Shanghai Astronomical Observatory, Shanghai 200030, China                \and
           Department of Physics, Massachusetts Institute of Technology, Cambridge, MA 02139, USA \and
            Graduate Institute of Astrophysics and Department of Physics, National Taiwan University, Taipei 10617, Taiwan \and
             Department of Physics, National Chung Hsing University, Taichung 402, Taiwan\\
	     }

\authorrunning{Lin et al.}

   \date{Received XX; accepted XX}

 
\abstract
  {The halo assembly bias, a phenomenon referring to dependencies of the large-scale bias of a dark matter halo other than its mass, is a fundamental property of the standard cosmological model.  
  First discovered in 2005 from the {\it Millennium Run} simulation,
it has been proven {\it very} difficult to be detected observationally, with only a few convincing claims of detection so far. The main obstacle lies in finding an accurate proxy of the halo formation time.
In this study, by utilizing a constrained simulation that can faithfully reproduce the observed structures larger than $2\,$Mpc in the local universe, for a sample of  634 massive clusters at $z\le 0.12$, we find their counterpart halos in the simulation and use the mass growth history of the matched halos to estimate the formation time of the observed clusters. 
This allows us to construct a pair of early- and late-forming clusters, with similar mass as measured via weak 
gravitational lensing, and large-scale bias differing at  
$\approx 3\sigma$ level,   
suggestive of
the signature of assembly bias, 
which is further corroborated by the properties of cluster  galaxies, including the brightest cluster galaxy, and
the spatial distribution and number of member galaxies.
Our study paves a way to further detect assembly bias based on cluster samples constructed purely on observed quantities.
}


\maketitle
%

\section{Introduction} 
\label{sec:intro}

It has been well-known for more than a decade that, although the large-scale bias of dark matter halos is primarily a function of halo mass, 
it also has secondary dependencies on other halo properties such as the formation time, concentration, spin, etc \citep[e.g.,][]{gao05,jing07}.  Such  dependencies are loosely referred to as assembly bias (AB; see e.g., \citealt{mao18,wechsler18}).
As AB is a subtle, yet solid feature of 
the cold dark matter model with a cosmological constant ($\Lambda$CDM; e.g., \citealt{contreras21}), a solid observational detection of the phenomenon will serve as a critical validation of the  model.

In this study, we restrict ourselves to investigating the AB as manifested by the differences in  halo formation time; that is, we aim to detect  differences in the large-scale bias for halos of the same mass but with different formation times.
Numerous studies have shown that the amplitude of halo AB is dependent on the halo mass
 (and to some degree, the definition of halo formation time and even the definition of a halo itself; see \citealt{mansfield20} for details).  
In general it is more prominent for low mass halos, while less so at the  massive end  such as galaxy clusters.
In our earlier attempt to detect AB in halos of mass comparable to that of the Milky Way ($\sim 10^{12}\,M_\odot$), we did not find convincing evidence for AB \citep{lin16}.  It is likely due to our proxy for halo formation time, namely the mean stellar age of central galaxies derived from spectra provided by the Sloan Digital Sky Survey (SDSS; \citealt{york00}), is not of sufficient accuracy.

Using a large sample of SDSS redMaPPer clusters \citep{rykoff14}, \citet[][hereafter M16]{miyatake16} claimed a strong detection of AB, which unfortunately turned out to be mainly due to the projection effect; briefly speaking, the formation time proxy used by M16, namely the spatial concentration of photometrically selected potential cluster member galaxies, is  contaminated by large-scale correlated structures along the line-of-sight, which mimics the AB signal \citep{more16,zu17,sunayama19}.

It is thus clear that, in the pursuit of detection of the AB signal, one of the main challenges is to have a robust proxy of halo formation time (while ensuring similar halo masses between early- and late-forming samples  being another challenging aspect).

Here we present a novel approach to the estimation of halo formation time, which makes heavy use of the forward-modeling-based numerical simulation {\it Elucid} \citep{wang16}. {\it Elucid} is designed to reproduce the structures larger than $\sim 2\,$Mpc as observed by the SDSS main galaxy sample \citep{strauss02} out to a redshift $z=0.12$.
As such, by matching a galaxy cluster sample drawn from the cluster and group catalog of \citet[][hereafter Y07]{yang07}, we can find a one-to-one correspondence between the observed clusters and the simulated halos, whereby the cluster formation time is derived from the halo mass growth history.  This method then allows us to split the cluster sample into early- and late-forming subsamples; based on mass measurements from weak gravitational lensing (WL) and large-scale bias derived from cluster-galaxy cross correlation, we show that,
within the $\Lambda$CDM framework,
 a pair of early- and late-forming cluster samples exhibits the signature of AB at $\gtrsim 3\sigma$ level, which will facilitate the study of halo AB at the high mass end (see also \citealt{zu21}).

This paper is structured as follows: in Section~\ref{sec:method} we describe the key elements of our analysis.  We construct a pair of early- and late-forming cluster samples that exhibits the AB signal, and examine the properties of cluster galaxy population of the samples in
Section~\ref{sec:res}.  
We discuss the validity of our approach in Section~\ref{sec:foundation}, and
implications and prospects of the method developed in this paper in Section~\ref{sec:disc}.
Throughout this paper we adopt a {\it WMAP5}  \citep{komatsu09} $\Lambda$CDM  model,
where $\Omega_m=0.258$, $\Omega_\Lambda=0.742$, $H_0=100h~{\rm km\,s^{-1}\,Mpc^{-1}}$ with $h=0.71$, $\sigma_8=0.8$, which is employed by {\it Elucid} \citep{wang14b,wang16}.
All optical magnitudes are in the AB photometry system \citep[][which should not to be confused with assembly bias]{oke83}.

\section{Methodology} 
\label{sec:method}

Here we provide 
an overview of the main elements of our analysis.  
The basis of this work, the constrained simulation {\it Elucid}, is presented  in Section~\ref{sec:elucid}.
Our cluster sample, and the way we match it to the simulated halos from {\it Elucid}, are described in Section~\ref{sec:sample}.
We then turn to our key observables, clustering and WL measurements, in Sections~\ref{sec:xc} and \ref{sec:wl}.
We describe our method for measuring the cluster galaxy surface density profiles in Section~\ref{sec:surfden}.

\subsection{Constrained Simulation {\it Elucid}} 
\label{sec:elucid}

The goal of producing the {\it Elucid} simulation is to reproduce the large-scale structures as observed by SDSS, which would then allow us to ``visualize'' the distribution of dark matter, and better understanding how galaxies populate dark matter halos.
The methodology behind {\it Elucid} can be found in \citet{wang14b}.  Basically, a nonlinear density field is provided to a Hamiltonian Markov Chain Monte Carlo (HMC) algorithm combined with particle-mesh (PM) dynamics, which is able to reconstruct the initial linear density field.  That density field is then evolved to the present-day with high resolution $N$-body simulations.

For the specific simulation used in this work, the nonlinear density field is constructed based on the group catalog of  Y07, which itself is based on SDSS data release 7 (DR7; \citealt{sdssdr7}).  
Given that the galactic systems in the Y07 catalog are complete down to a mass limit of $\approx 10^{12} h^{-1}\,M_\odot$ at $z\sim 0.12$, the nonlinear density field is estimated using only halos above that mass limit and in the redshift range $0.01-0.12$, lying within the northern Galactic cap region of the DR7 footprint.
The reconstructed initial density  field is then evolved  with $3072^3$ particles in a 500$h^{-1}$\,Mpc box, using a modified version of  GADGET-2 \citep{springel05,wang16}. 
Tests based on detailed mock galaxy samples  showed that the typical scatter between the true (nonlinear) density field and the reconstructed one is 0.23 dex
when smoothed over a scale of $2h^{-1}\,$Mpc \citep{wang16}, which is about the typical size of a massive cluster, and is thus a good match for our purpose.

\subsection{Galaxy Cluster Sample Selection} 
\label{sec:sample}

To match real clusters  to the simulated halos in {\it Elucid}, we use the model C version of the Y07 catalog.  As shown in \citet{wang16}, only part of the SDSS DR7 footprint is covered in the reconstructed volume, and we end up with 644 clusters with mass $M_{200m}\ge 10^{14}h^{-1}\,M_\odot$.  Here the mass is defined within a radius $r_{200m}$, within which the mean density is 200 times the mean density of the universe at the redshift of the cluster.
We note that the mass estimates of galactic systems in Y07 is based on a method similar in spirit to subhalo abundance matching \citep[e.g.,][]{conroy06,wechsler18}, that is, given the dark matter halo mass function, one assumes the total stellar mass (or luminosity) content of a galactic system is directly proportion to the halo mass, and thus one can ``assign'' a halo mass to observed groups/clusters to halos of the same spatial density.

The matching between the Y07 clusters and the {\it Elucid} halos is done in the following fashion: for a
given  cluster with mass $M_1$, we search all simulated halos with a distance to the
position of the cluster in the simulation less than $4 h^{-1}\,$Mpc,
which is the length scale adopted in \citet{wang16} for the density field reconstruction.
We have properly taken into account the redshift space distortion effect of clusters, by moving dark matter halos to redshift space using their peculiar velocities along the line of sight.
Let us denote the mass of halos that lie within the sphere as $M_2$. For all halos with mass satisfying $|\log(M_1 /M_2)| <0.5$,  we select the one  with
the smallest $|\log(M_1 / M_2)|$ as the matched halo. 
As demonstrated by \citet{wang16}, such criteria of matching enable $>95\%$ of clusters more massive than $10^{14}h^{-1}\,M_\odot$ to be associated with a halo in {\it Elucid}.
If no halo satisfies the above condition, then  
the cluster is discarded  from the sample.  
Out of 644 clusters, we find 634 matches this way.\footnote{ 
Choosing a smaller matching radius (e.g., $3h^{-1}\,$Mpc) and a more strict mass ratio constraint  (e.g., $|\log(M_1 /M_2)| <0.3$) will result in 540 matches.  For our main cluster samples (to be presented in Section~\ref{sec:z20ex}), the reduction of sample size is around 80\%, with very similar halo mass distributions, and thus will not change our conclusions.}

For the clusters with a counterpart halo, we extract the mass growth history of the main subhalo (that is, simply following the main trunk of the merger tree), and derive several formation time indicators, such as $z_{80}$, $z_{50}$,   $z_{20}$, and $z_{\rm mah}$.  While the first three 
 correspond to the redshifts when a halo first reaches 80\%, 50\% and 20\% of its final ($z=0$) mass, 
the last quantity is obtained by first fitting the mass growth history by the form
\begin{equation}
M(z) \propto \exp(-\alpha z),
\end{equation}
with the Levenberg-Marquardt algorithm for minimization of the least squares, 
and then setting $z_{\rm mah}\equiv 2/\alpha-1$ \citep{wechsler06}.

Finally, we note that the cluster center defined by the Y07 cluster finding algorithm is a luminosity weighted position based on the distribution of member galaxies, and the brightest cluster galaxies (BCGs) are not necessarily located at the very center.

\subsection{Cluster-Galaxy Cross Correlation Function} 
\label{sec:xc}

With only $\sim 600$ clusters, inferring the large-scale bias $b$ via auto-correlation function is   challenging and the result will be noisy.  We naturally opt for cluster-galaxy cross correlation $w_{\rm p,gc}$; 
by comparing the $w_{\rm p,gc}$ of early- and late-forming clusters at scales $10-30 h^{-1}\,$Mpc, we can then infer the relative bias  of the two populations of clusters.

Following \cite{guo17}, we measure the cross-correlation function $w_{\rm p}(r_p)$ between our cluster samples and a volume-limited galaxy sample drawn from SDSS DR7 main spectroscopic sample with the $r$-band absolute magnitude $M_r\le -20.5$ and $0.02\le z\le 0.132$. The $w_{\rm p}(r_p)$ measurement are calculated using the \citet{landy93} estimator, with $r_{\rm p}$ being the projected separation of the cluster-galaxy pairs.  We choose logarithmic $r_p$ bins with a width $\Delta\log r_p = 0.2$ from $0.1$ to $63.1 h^{-1}$Mpc.  The maximum line-of-sight integration length $\pi_{\rm max}$ 
is set to $40 h^{-1}$Mpc.
Setting $\pi_{\rm max}$ to larger values does not change our results.
The uncertainties in the $w_{\rm p}$ measurements are obtained by running 400 jackknife resampling.

\subsection{Weak Gravitational Lensing Measurements} 
\label{sec:wl}

We measure weak gravitational lensing signals as the average excess surface mass density $\Delta\Sigma$ by stacking $\sim600$ clusters in the same manner as M16, which followed the procedure described in \citet{mandelbaum13}. We use the shape catalog based on the photometric galaxy catalog from SDSS DR8 \citep{reyes12}. 
The shapes of source galaxies were measured by the re-Gaussianization technique \citep{hirata03}. Systematic uncertainties in the shape measurements were investigated as done in \citet{mandelbaum05} and calibrations were performed using image simulations from \citet{mandelbaum12}.
Their photometric redshifts (photo-$z$) are estimated using the publicly available  code ZEBRA \citep{feldmann06, nakajima12}. We choose logarithmic bins $r_p$ with a width $\Delta \log r_p = 0.14$ from 0.025 to $50 h^{-1}$Mpc. We apply the photo-$z$ correction, boost factor correction, and random signal correction following \citet{mandelbaum05} and \citet{nakajima12}. We estimate the covariance matrix using the jackknife technique as described in M16.

 To infer the cluster mass, we fit a halo model that is similar to what is described in M16. We fit the signal within the range of $0.3 < r_p/[h^{-1}\mathrm{Mpc}] < 3$, since this scale is not affected by 2-halo term. 
We use a truncated \citet[][hereafter NFW]{navarro97} profile, described in \citet{takada03,takada03b}, 
  and assume that there are some fraction of off-centered clusters with respect to their true center. There are four fitting parameters in total: cluster mass $M_{200m}$, concentration parameter $c_{200m}$, fraction of centered clusters $q_\mathrm{cen}$, and typical off-centering scale with respect to $r_{200m}$, $\alpha_{\rm cen}$. For the $q_\mathrm{cen}$, we employ a Gaussian prior $\mathcal{N}(0.8, 0.1)$.
The cluster mass constraints are insensitive to the choice of the prior; when adopting a prior $\mathcal{N}(0.8, 0,2)$, the change in cluster mass is typical only a few percent.

\subsection{Cluster Galaxy Surface Density Measurements} 
\label{sec:surfden}

To probe the properties of member galaxies  of the clusters, we cross-correlate the clusters with photometric galaxies detected in  SDSS. This cross-correlation technique makes use of the fact that members  associated with the clusters will introduce a galaxy overdensity  along the line-of-sight. By measuring the average number density of interlopers via random sightlines and subtracting such a component from the average number density of galaxies around the clusters, one can probe the number density of member galaxies and their observed properties statistically
(see e.g., \citealt{lin04,wang12b,lan16,tinker21}).

In practice,
we first select robustly detected photometric galaxies 
with $z$-band apparent magnitudes $m_z\le 21$
and estimate the absolute magnitude of a galaxy around a cluster at the redshift of the cluster with a $k$-correction. We adopt the same method as \citet{lan16} to obtain the $k$-correction, 
by using the median $k$-correction of SDSS spectroscopic galaxies from \citet{blanton05} with similar observed $(u-r)$ and $(g-r)$  colors. We further separate galaxies into blue and red populations using the color-magnitude relation from \citet{baldry04} and only include galaxies with $M_r\le -18$, which is the
completeness limit for both types of galaxy populations. Finally, we count the numbers of blue and red galaxies around the clusters as a function of projected distance and estimate the number density of interlopers with 10 random sightlines for each cluster. 
Unlike our $w_p$ and WL measurements, here we use the BCG as the cluster center.
The  uncertainty is obtained by bootstrapping the samples 500 times.

\section{Results} 
\label{sec:res}

\subsection{Cluster Selection using $z_{50}$ and $z_{\rm mah}$}

\begin{figure}
\vspace{-10mm}	
	\includegraphics[width=\columnwidth]{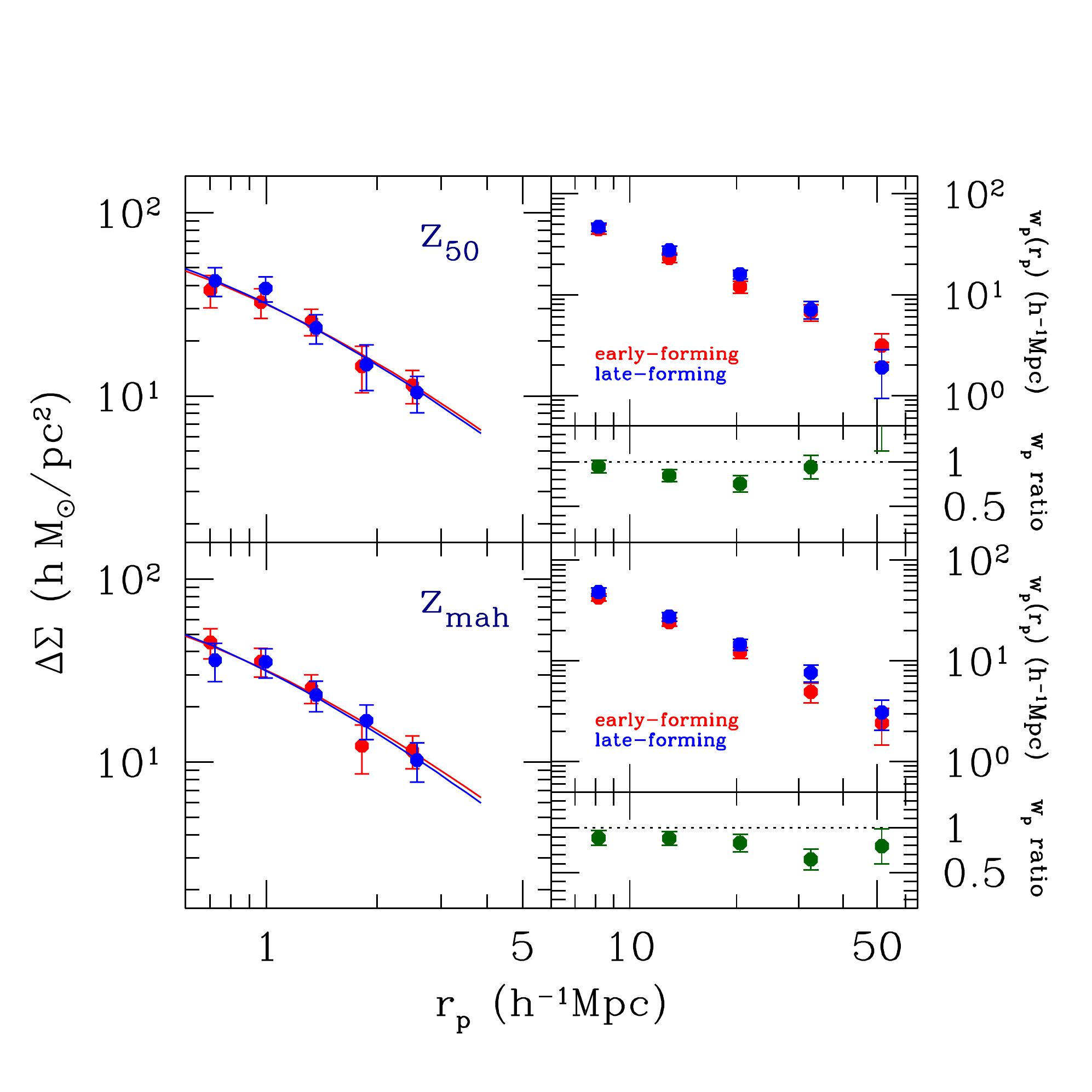}
\vspace{-10mm}	
    \caption{Weak lensing (more specifically, the excess surface mass density, left) and clustering (right) measurements as a function of projected distance for clusters split in half by $z_{50}$  (upper panels) and $z_{\rm mah}$ (lower panels).  
The red and blue symbols represent early- and late-forming clusters, respectively.  There are 323 and 311 early- and late-forming clusters when split by $z_{\rm 50,div}=0.521$, while 316 and 318 clusters when split by $z_{\rm mah,div}=0.469$.  WL masses for these samples are all very close, about $1.5\times 10^{14}h^{-1}\,M_\odot$ 
(the curves in the left panels show the best-fit models). 
The green points in the lower right panel are the ratio of the early-to-late $w_{\rm p}$ measurements, which is equivalent to the large-scale bias ratio $b_{\rm early}/b_{\rm late}$.  
The signature of AB at cluster scales is manifested by $b_{\rm early}/b_{\rm late}<1$, which is consistent with our measurements.
The probability for these samples to be drawn from the same parent sample is $p=0.0258$ (for the $z_{50}$-selected samples) and $p=0.0295$ (for the $z_{\rm mah}$-selected ones). }
\label{fig:z50}
\end{figure}

\begin{figure*}
\vspace{-5mm}	
\hspace{20mm}
	\includegraphics[width=1.6\columnwidth]{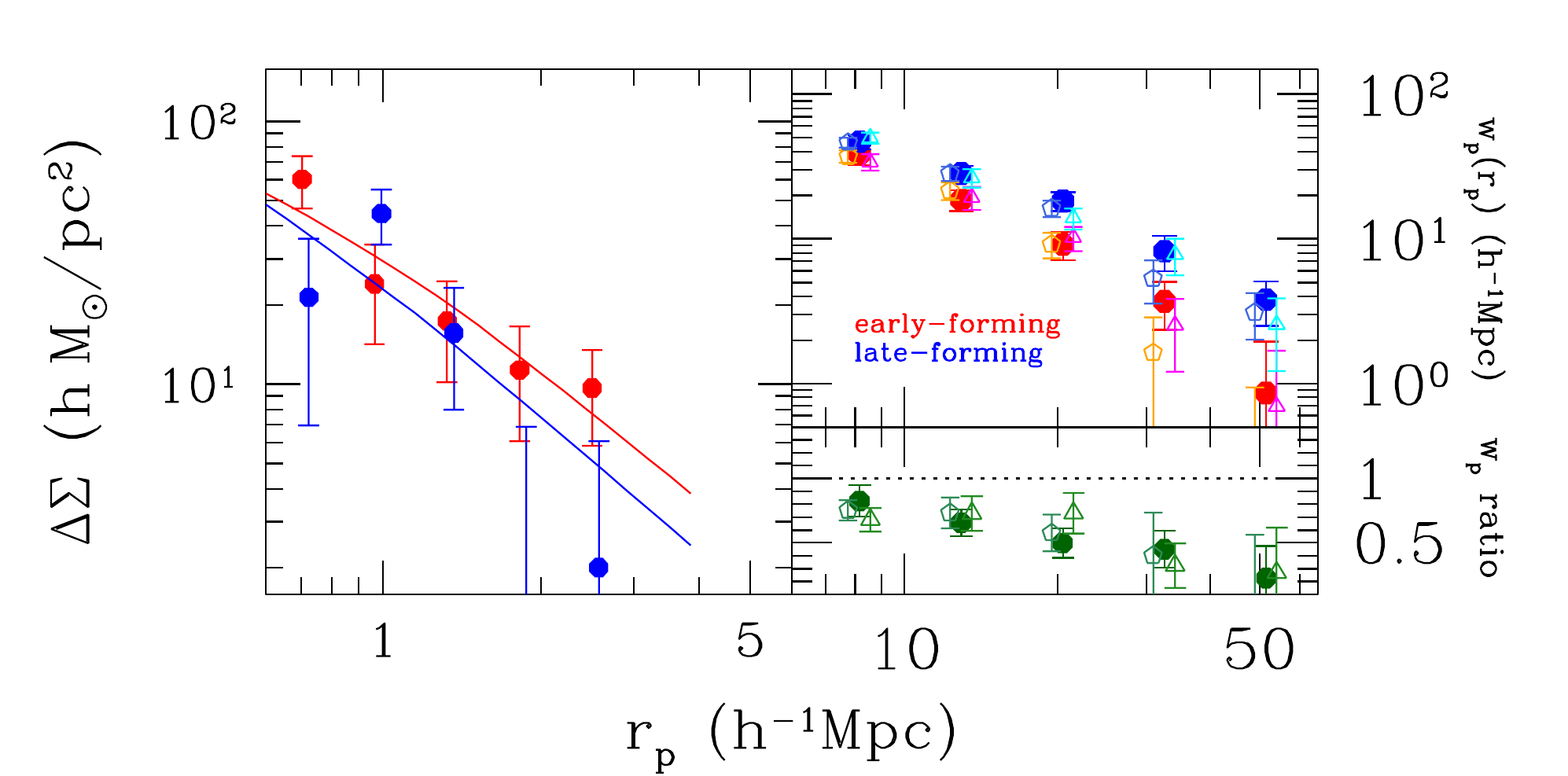}
    \caption{
    Similar to Fig.~\ref{fig:z50}, but for the
$z_{\rm 20ex,Mz}$ set, which is  selected to lie within the parameter space $\log M_{200m}/(h^{-1}\,M_\odot)=14-14.5$ and $z=0.06-0.12$.
It consists of 138 oldest (with $z_{20}>1.35$; red solid points) and 121 youngest (with $z_{20}<0.85$; blue solid points), with masses within $1\sigma$ of each other (the curves in the left panel show the best-fit models); the probability for the two cluster samples to have the same large-scale bias is $p=1.03\times 10^{-10}$ using the clustering measurements in the mid-range.  
In the upper right panel we show the projected correlation function of the {\it Elucid} halos that correspond to the clusters in the $z_{\rm 20ex,Mz}$ set (namely the ``counterpart halos'') as open pentagons, and that of the early- and late-forming {\it Elucid} halos selected from nearly the same halo mass range to have identical mean masses ($\approx 1.25\times 10^{14} h^{-1}\,M_\odot$; the ``equal mass halos'' -- open triangles).  While the clustering amplitudes of the latter are similar to the $z_{\rm 20ex,Mz}$ set, those of the  counterpart halos deviate from the real clusters at scales larger than $30 h^{-1}\,$Mpc.
The early-to-late forming relative biases of the $z_{\rm 20ex,Mz}$ set, the counterpart halos, and the equal mass halos, are shown in the lower right panel as the green solid points, open pentagons, and open triangles, respectively.
For both types of open symbols in the right panels, the horizontal locations are slightly offset from the actual values, for the sake of clarity.  The error bars are obtained from the covariance matrix constructed from  400 jackknife resamples for all our clustering measurements.
}
\label{fig:z20mz}
\end{figure*}

We start by splitting the clusters by by either $z_{50}$ or $z_{\rm mah}$,  using $z_{\rm 50,div}=0.521$ and $z_{\rm mah,div}=0.469$ for separating early- and late-forming clusters; the division redshifts are  chosen to make the numbers of clusters in the two samples as close as possible.  The mean masses of the resulting $z_{50}$-selected 323 early-forming and 311 late-forming clusters, from stacked WL, are $M_{200m}=(1.54\pm 0.22)\times 10^{14}h^{-1}\,M_\odot$ and $(1.51\pm 0.22)\times 10^{14}h^{-1}\,M_\odot$, respectively.  In Figure~\ref{fig:z50} (upper panels), we show the lensing signals (i.e., the surface mass density contrast) on the left side, while the projected cluster-galaxy correlation function  on the upper right panel.  
At cluster scales, it is expected that late-forming clusters would have higher clustering amplitude (i.e., larger large-scale bias) compared to the early-forming ones due to AB, which is consistent with our measurements.
The lower right panel shows the ratio of  $w_{\rm p,early}/w_{\rm p,late}=b_{\rm early}/b_{\rm late}$.  
To quantify the difference  between the $w_{\rm p,early}$ and $w_{\rm p,late}$ measurements (in terms of their ratio), we follow the methodology developed in \citet[][see Section 4.1.2 therein]{lin16} and calculate\\
\begin{equation}
\chi^2 = \sum_{ij} \left( w_{\rm e}(r_i) /w_{\rm l}(r_i)  - \xi  \right) C_{ij}^{-1} \left(  w_{\rm e}(r_j) / w_{\rm l}(r_j) -\xi  \right),
\label{eq:chi2}
\end{equation}
where $w_{\rm e}$ and $w_{\rm l}$ are shorthands for $w_{\rm p,early}$ and $w_{\rm p,late}$, respectively,  $C$ is the covariance matrix built from the ratio between $w_{\rm p,early}$ and $w_{\rm p,late}$ and their associated jackknife samples, and constant $\xi=1$.  
The errorbars of the $w_{\rm p}$ ratio shown in Figure~\ref{fig:z50} (as well as in Figures~\ref{fig:z20mz} and \ref{fig:20p}) are calculated based on the diagonal terms of the covariance matrix $C$.
With $\chi^2=9.3$ from 3 degrees of freedom, over the scales from $12.9 h^{-1}\,$Mpc to $32.5 h^{-1}\,$Mpc (hereafter the ``mid-range''),
we find that the two samples have a probability $p=0.0258$  to be drawn from the same parent population (i.e., having the same large-scale bias).  Using the full scale as shown in the Figure ($8.2 h^{-1}\,$Mpc to $51.5 h^{-1}\,$Mpc), the probability changes to $p=0.0393$.

A similar result using cluster samples defined by $z_{\rm mah}$ is shown in the lower panels of Figure~\ref{fig:z50}.   The mean masses of 316 early-forming  and 318 late-forming clusters  are $M_{200m}=(1.52\pm 0.23)\times 10^{14}h^{-1}\,M_\odot$ and $(1.46\pm 0.21)\times 10^{14}h^{-1}\,M_\odot$, respectively.  
Using Eqn.~\ref{eq:chi2}, we find that the two samples have  a probability $p=0.0295$ ($0.0152$)  to be consistent using the measurements from the mid-range (full-scale).

We then further seek stronger signals by exploring the extremal of the age distribution.
However, as pointed out by \citet{chue18}, at the mass scale of our clusters (i.e., around $10^{14} h^{-1}\,M_\odot$), $z_{50}$ may not be the best age indicator.  Quantities such as $z_{20}$ or $z_{80}$ may better reflect the formation history.

\subsection{Age Extremum Selection of Clusters} 
\label{sec:z20ex}

\begin{figure}
\vspace{-5mm}	
	\includegraphics[width=\columnwidth]{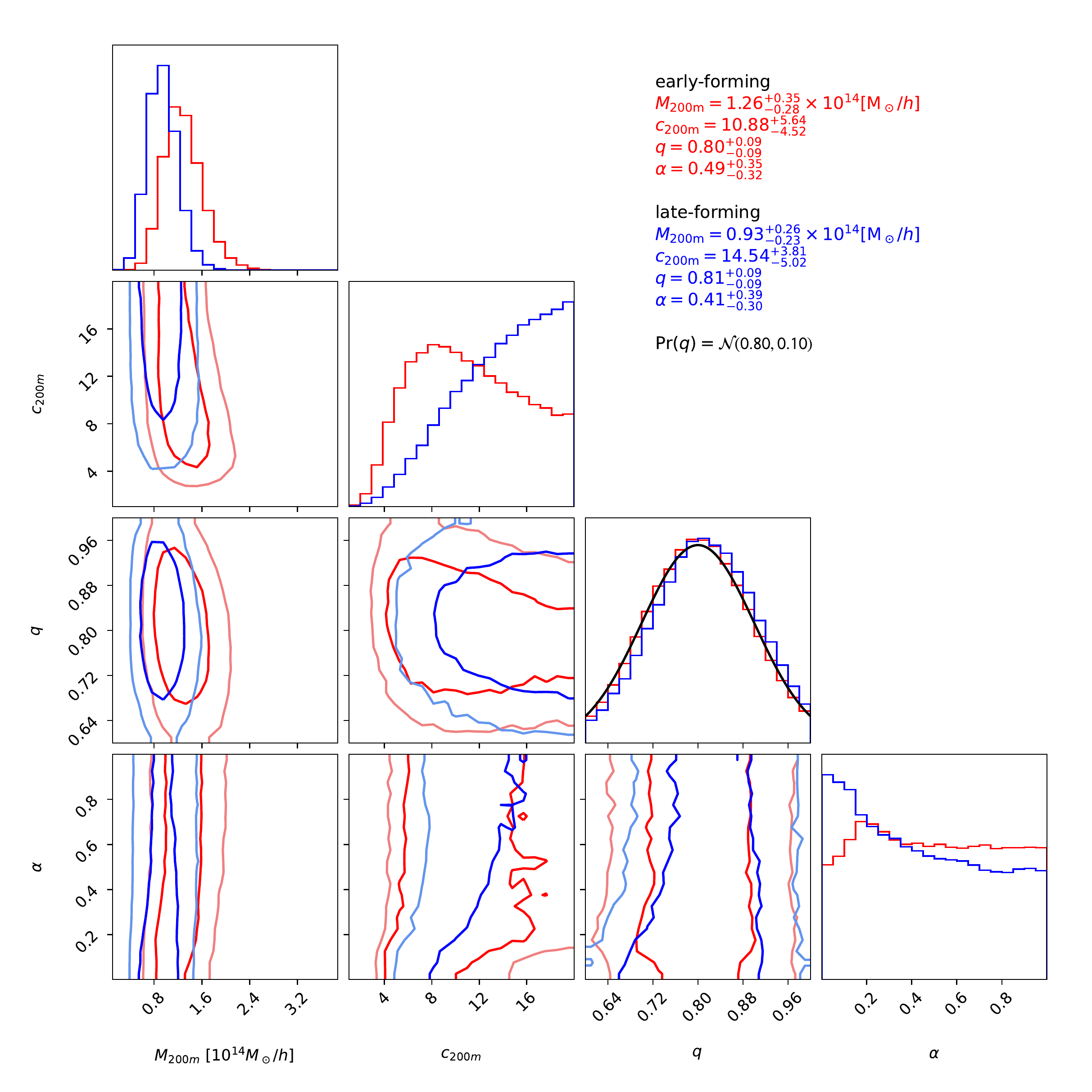}
\vspace{-7mm}		
    \caption{
Corner plot showing $1\sigma$ \& $2\sigma$ (darker and lighter colored, respectively) confidence levels of our measurements and the nuisance parameters ($q_{\rm cen}$ \& $\alpha_{\rm cen}$; see Section~\ref{sec:wl}) for the $z_{\rm 20ex,Mz}$ set.  
While we can confidently constrain the cluster masses (e.g., top left panel), the constraining power on total mass concentration is unfortunately weak (top panel in the second column from the left).     
}
\label{fig:corner}
\end{figure}

After testing with various age indicators mentioned above, 
it is found that, 
by selecting 138 oldest clusters  with $z_{20}>1.35$, paired with 121 youngest clusters with $z_{20}<0.85$, both limited in the redshift range $z=0.06-0.12$ and mass range $\log M_{200m}/(h^{-1}\,M_\odot)=14-14.5$ (as estimated by Y07),   an   AB-like signal can be seen. 
 The mean masses based on WL are measured to be  $M_{\rm 200m,early}=1.26^{+0.35}_{-0.29} \times 10^{14} h^{-1}\,M_\odot$ and $M_{\rm 200m,late}=0.95^{+0.26}_{-0.22} \times 10^{14} h^{-1}\,M_\odot$, consistent within $1\sigma$ (Figures~\ref{fig:z20mz} \& \ref{fig:corner}).  
While the dividing redshifts are chosen so that we have at least 120 clusters in each sample,  the redshift and halo mass ranges are selected to facilitate consistent cluster mass estimates.  
We shall refer to this pair of cluster samples as the $z_{\rm 20ex,Mz}$ set.
We note in passing that the mean masses based on Y07 are $(1.51\pm 0.50)\times 10^{14}h^{-1}\,M_\odot$ and $(1.66\pm 0.55)\times 10^{14}h^{-1}\,M_\odot$ for the early- and late-forming samples, respectively.
In Tables~\ref{tab:early} \& \ref{tab:late} we provide some basic information of the early- and late-forming cluster samples, respectively, which include cluster ID, cluster mass based on Y07, cluster center taken from the Y07 catalog, redshift, and $z_{20}$.
Given the mass difference between the early- and late-forming clusters, we  should compare the measured $b_{\rm early}/b_{\rm late}$ to the theoretically expected bias ratio $r_{\rm b,th}=1.11$ (assuming no effects from AB, as obtained by the bias--halo mass relation of \citealt{tinker10}), in order to evaluate the probability that the two samples have the same large-scale bias.
We thus set $\xi = r_{\rm b,th}$ in Eqn.~\ref{eq:chi2}, 
and find the probabilities to be $p=1.02\times 10^{-10}$ and $1.16\times 10^{-11}$ using the data from the mid-range and full scale, respectively.

We note that the mean masses of the {\it Elucid} {\it halos} that correspond to the early- and late-forming {\it clusters} are $(1.36 \pm 0.69)\times 10^{14} h^{-1}\,M_\odot$ and $(1.62 \pm 0.77)\times 10^{14} h^{-1}\,M_\odot$, respectively.  
We shall refer to these halos as the ``counterpart'' halos.  
While the WL mass of the early-forming cluster sample is close to the mean mass of the counterpart halos, the situation is  different for the late-forming clusters and their associated halos.  Not only do the masses differ (at $1.5-2\sigma$ level), the sign is also changed between the observed and simulated samples (in the $z_{\rm 20ex,Mz}$ set, the late-forming cluster sample has a lower mean mass than the early-forming ones, opposite to the simulated halos).
We show the ratio of projected cross-correlation function (PCCF) of early-to-late-forming halos as open pentagons in the lower right panel of Fig.~\ref{fig:z20mz}.  Although the ratios are similar to that of the $z_{\rm 20ex,Mz}$ set, we note that the actual amplitude of clustering (i.e., the absolute values of bias) of the counterpart halos differ somewhat from the real clusters at scales larger than $30 h^{-1}\,$Mpc (see the open pentagons in the upper right panel).
The PCCFs of the counterpart halos are obtained by cross correlating these halos with  
low mass halos in the mass range $\log M_{200m}/(h^{-1}M_\odot)=11.5-12.5.$\footnote{With coordinates of all halos transformed from the ($x$, $y$, $z$) space of {\it Elucid} to (RA, Dec., redshift) space first.}
To make an accurate comparison of the resulting $w_{\rm p,early,co}$ \& $w_{\rm p,late,co}$ of the counterpart halos with those of the $z_{\rm 20ex,Mz}$ set, we also cross correlate the observed clusters with the same set of low mass halos in {\it Elucid}, and multiply $w_{\rm p,early,co}$ \& $w_{\rm p,late,co}$ with a factor $\zeta$ that makes the amplitudes of the $z_{\rm 20ex,Mz}$ cluster--halo PCCF and the $z_{\rm 20ex,Mz}$ cluster--SDSS galaxy PCCF at $8.2 h^{-1}\,$Mpc identical.

Given the relatively large uncertainties in the WL mass measurements of the $z_{\rm 20ex,Mz}$ set, we have to test their effects on the AB signal exhibited by our cluster samples, by considering an extreme case where the differences in the large-scale clustering are {\it maximally} due to the cluster mass difference.  
To do so, we consider two cases: (1) assuming that the mean mass of the early-forming cluster sample is biased high by $1\sigma$, while that of the late-forming one is biased low by $1\sigma$, and (2) assuming that the mean mass of the late-forming sample is biased low by at least $2\sigma$.
While there is a 2.6\% probability for the first case to occur, the likelihood for the second case is 2.3\%.
For the first case, 
we set $\xi = r_{\rm b,th}$ to  0.92,  the value corresponding to the bias ratio of halos of masses $M_{\rm 200m,early,min}=0.97\times 10^{14} h^{-1}\,M_\odot$ (i.e., $1\sigma$ lower than the mean mass of the early-forming clusters) and $M_{\rm 200m,late,max}=1.21\times 10^{14} h^{-1}\,M_\odot$ ($1\sigma$ higher than the mean mass of the late-forming clusters) in Eqn.~\ref{eq:chi2}.  
In the  calculation above, we have again used  the bias--halo mass relation of \citet{tinker10}, and set $\xi = r_{\rm b,th}$ in Eqn.~\ref{eq:chi2}.
Using the mid-range (full scale) clustering measurements, we find that
the probability for the pair of cluster samples to have the same bias becomes $p=7.49\times 10^{-5}$ ($5.40\times 10^{-5}$). 
As for the second case, we keep the mass of the early-forming sample at the measured value, while increase the mean mass of the late-forming one to $1.47\times 10^{14}h^{-1}\,M_\odot$.  The expected bias ratio is $ r_{\rm b,th}=0.94$, which is high compared to our measurements.  The probabilities are $p=2.44\times 10^{-5}$ and $1.35\times 10^{-5}$ using the mid-range and full scale data, respectively. 

However, if we take into account the potential uncertainties of  in the theoretical predictions of the bias ratio from \citet{tinker10}, and artificially decrease $ r_{\rm b,th}$ by 10\%, the probabilities become 
$5.30\times 10^{-3}$ and $4.53\times 10^{-3}$ (for case 1)
$2.50\times 10^{-3}$ and $2.13\times 10^{-3}$ (for case 2) using the mid-range and full scale data, respectively; 
these correspond to significance levels of $2.8\sigma$ (case 1) and $3\sigma$ (case 2).

In the upper (lower) right panel of Figure~\ref{fig:z20mz}, we further show the theoretically expected PCCFs ($b_{\rm early}/b_{\rm late}$ ratio) as the open triangles, as derived from {\it Elucid}, in which the effect of AB {\it must} be present.  We first 
select dark matter halos with masses $M_{200m}=(0.7-1.9)\times 10^{14} h^{-1}\,M_\odot$, then separate them into early- and late-forming using the same redshift division as done for the $z_{\rm 20ex,Mz}$ set.  
This pair of halos will be referred to as the ``equal mass halos''.
We then cross-correlate the resulting 98 early-forming and 82 late-forming halos, which have mean masses of $(1.24\pm 0.31)\times 10^{14} h^{-1}\,M_\odot$ and $(1.26\pm 0.36)\times 10^{14} h^{-1}\,M_\odot$, respectively, with 
low mass halos in the mass range $\log M_{200m}/(h^{-1}M_\odot)=11.5-12.5$,
and derive the ratio between $w_{\rm p,early,eq}$ and $w_{\rm p,late,eq}$.  
The triangles appear to be  consistent with our measurements (the solid points), indicating that the  
AB signal from the $z_{\rm 20ex,Mz}$ set is similar in amplitude to that expected in $\Lambda$CDM.
The probability for the early- and late-forming {\it halos} to have the same large-scale bias  is $p=1.78\times 10^{-4}$ using the mid-range clustering measurements.
Similar to the PCCFs  of the counterpart halos, the PCCFs of the equal mass halos (shown as open triangles in the upper right panel of Fig.~\ref{fig:z20mz}) have been adjusted in amplitude by the same factor $\zeta$.

We conclude by noting that, while the mean masses and mass ranges of the ``counterpart'' halos and ``equal mass'' halos differ, both exhibit a similar degree of AB.  Therefore, despite the WL mass uncertainties of the $z_{\rm 20ex,Mz}$ set, 
if AB exists in the real Universe at the predicted amplitude, the signal should still be detectable.

\begin{table*}
	\centering
	\caption{Early-Forming Clusters of the $z_{\rm 20ex,Mz}$ set (the full table is available online)}
	\label{tab:early}
	\begin{tabular}{rrrrcc} 
		\hline
		ID & $\log M_{200m}$ & R.A. & Dec. & $z$ & $z_{20}$\\
		    &  ($h^{-1} M_\odot$) & (J2000)  & (J2000)  & & \\
		\hline
87 & 14.494600 & 257.446899 & 34.438801 & 0.085400 & 1.383961\\
90 & 14.484900 & 132.532898 & 29.548309 & 0.104570 & 1.456453\\
99 & 14.464800 & 227.795395 & 5.274312 & 0.080080 & 1.531159\\
101 & 14.459500 & 179.268997 & 5.061248 & 0.074340 & 1.531159\\
103 & 14.456200 & 230.749695 & 30.984060 & 0.112770 & 1.608133\\
		\hline
	\end{tabular}
\end{table*}

\begin{table*}
	\centering
	\caption{Late-Forming Clusters of the $z_{\rm 20ex,Mz}$ set (the full table is available online)}
	\label{tab:late}
	\begin{tabular}{rrrrcc} 
		\hline
		ID & $\log M_{200m}$ & R.A. & Dec. & $z$ & $z_{20}$\\
		    &  ($h^{-1} M_\odot$) & (J2000)  & (J2000)  & & \\
		\hline
86 & 14.496500 & 227.337402 & 7.651995 & 0.077560 & 0.614971\\
89 & 14.486800 & 187.102997 & 12.065190 & 0.089030 & 0.476161\\
93 & 14.479200 & 134.566299 & 38.548759 & 0.093160 & 0.766834\\
97 & 14.468300 & 208.025497 & 46.378010 & 0.062580 & 0.714689\\
102 & 14.457900 & 186.548996 & 31.200930 & 0.060480 & 0.766834\\
		\hline
	\end{tabular}
\end{table*}

\subsection{Galaxy Properties of the $z_{\rm 20ex,Mz}$ set} 
\label{sec:gal}

We next examine various properties related to cluster galaxy populations in the $z_{\rm 20ex,Mz}$ set.
We first
measure the stacked galaxy surface density profiles of the early- and late-forming clusters in the $z_{\rm 20ex,Mz}$ set.  
Fitting the density profiles of red galaxies with $r$-band absolute magnitudes of $-24\le M_r \le -18$ with an NFW model, we find   concentration values of
$c_{\rm early,red}=7.1^{+1.7}_{-1.3}$ (with reduced $\chi_\nu^2=1.00$ with 16 degrees of freedom) and $c_{\rm late,red}=5.6_{-0.5}^{+0.6}$ ($\chi_\nu^2=0.74$),
which are consistent with the expectation that early-forming clusters would have a more spatially concentrated galaxy population, a premise of the analysis of M16.
Despite having a $33\%$ higher mean cluster mass, the early-forming clusters are found to have  25\% fewer  galaxies (of all colors) than the late-forming ones.  Given the halo occupation number measured from a sample of nearby clusters, one expects the cluster galaxy number $N \propto M_{200m}^{0.8}$ \citep[][]{lin04}; this makes the ``deficit'' of galaxies in our early-forming sample to be about 60\% compared to the case when AB does not play a role. 
Such a reduction in member galaxy number could be due to that, for an older cluster population, cluster galaxies are subject to dynamical processes such as tidal stripping for a longer time and can lose mass (and fade in luminosity as they age) and fall below our detection limit (e.g., \citealt{zentner05}).  In addition, there is also longer time for dynamical friction to drive more galaxies to the center and merge with the BCG.

One rough way to examine whether a cluster is relaxed or not is to check the proximity of its BCG to the center.  In the Y07 catalog, the cluster center is a luminosity weighted mean from all member galaxies.  We have calculated the  distances of BCGs from the cluster center for the  $z_{\rm 20ex,Mz}$ set, finding that BCGs in the early- and late-forming samples have a median separation from the center of $d_{\rm e}=(0.11\pm 0.01) r_{200m}$ and $d_{\rm l}=(0.14 \pm 0.01) r_{200m}$, respectively.  Such off-center values are consistent with our prior employed in the WL analysis (Section~\ref{sec:wl}), and are consistent with the expectation that older clusters would have BCGs closer to their center.

We further examine the formation time of the cluster galaxy properties, particularly the BCGs, using results from the full spectral fitting code {\tt STARLIGHT} \citep{cidfernandes13} as well as the spectral energy distribution  fitting technique presented in \citet{chang15}.  We  do not find any appreciable differences in the ages of member galaxies in the early- and late-forming clusters,  which might be due to the combination of (1) small cluster sample size, (2) insufficient depth of SDSS photometry and spectra, and the theoretical expectation that cluster galaxies have formed most of their stars prior to becoming members of the clusters we observe (e.g., \citealt{guo11};  for old stellar populations, it is extremely difficult to detect any age differences using tools currently available).

Finally, we investigate the magnitude gap $\Delta_{12}$ of the $z_{\rm 20ex,Mz}$ set, which is the differences in the $r$-band absolute magnitudes between the BCG and the second most luminous galaxy \citep{tremaine77}.
Additionally, we also examine $\Delta_{14}$,  the magnitude difference between the BCG and the fourth most luminous galaxy, as it has been suggested to be a more robust measure of the gap  \citep{golden-marx18}.
Using the cluster member catalog of Y07, we find that the median $\Delta_{12}=0.44\pm 0.01$ ($\Delta_{14}=0.99\pm 0.01$) with a scatter of 0.05 for the early-forming clusters, while that of the late-forming clusters is $\Delta_{12}=0.38\pm 0.01$ ($\Delta_{14}=0.87\pm 0.01$) with a scatter of 0.05.  
These results are again consistent with the expectation that the gap increases as a cluster ages, as more and more massive satellites are cannibalized by the BCG via dynamical friction \citep{ostriker75}.

To summarize, for the pair of cluster samples, we find from the properties of cluster galaxy populations evidence supporting their age differences, including (1) higher concentration of spatial distribution of red galaxies, 
(2) significantly reduced total galaxy number, (3) smaller offset of BCGs from cluster center, and (4)  larger magnitude gap, when comparing early-forming clusters with the late-forming ones.

\subsection{Null Tests} 
\label{sec:robust}

\begin{figure}
\vspace{-5mm}	
	\includegraphics[width=\columnwidth]{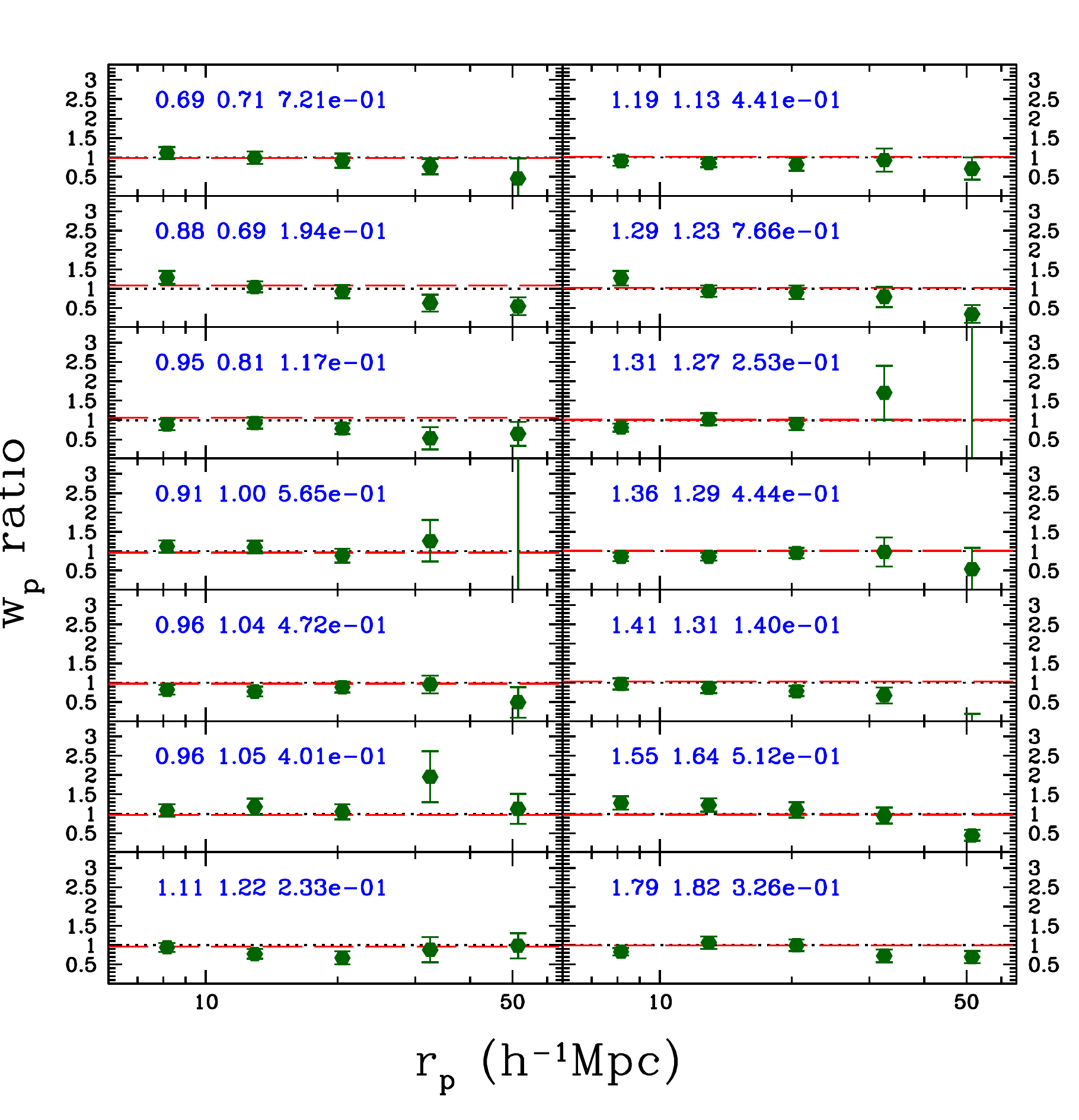}
\vspace{-7mm}		
    \caption{Ratio of $w_{\rm p,early}/w_{\rm p,late} = b_{\rm early}/b_{\rm late}$ of the 14 control samples.  The pairs are arranged in increasing masses (from top left to bottom left, continuing from top right to bottom right).  In each panel, the red horizontal dashed line denotes the theoretically expected bias ratio $r_{\rm b,th}$ (in the absence of AB); the 3 numbers shown in blue are the masses of the early-analogous and late-analogous clusters (in unit of $10^{14} h^{-1}\,M_\odot$), and the probability for the pair to be drawn from the same parent population. 
}
\label{fig:20p}
\end{figure}

As a sanity check, we construct 14 pairs of  ``control'' cluster samples that have similar distributions in halo mass and redshift as the $z_{\rm 20ex,Mz}$ set, but chosen from the parent cluster sample {\it regardless} of their $z_{20}$,
to demonstrate the robustness of our 
method.  
These samples are generated in the following way (and will be referred to as the control samples hereafter). 
We first make  grids in the $M$--$z$ space limited to that spanned by the $z_{\rm 20ex,Mz}$ set, that is, $\log M_{200m}/(h^{-1}\,M_\odot)=14-14.5$ and $z=0.06-0.12$.  To generate a control sample analogous to the early-forming clusters of the $z_{\rm 20ex,Mz}$ set, we only select clusters from grids in which the number of clusters from the full sample (of 634 clusters) is at least more than two larger than the number of early-forming clusters from the $z_{\rm 20ex,Mz}$ set.  A late-forming cluster sample analog is generated in a similar fashion. 
Note that the early-analog and late-analog samples are generated independently of each other.

In Figure~\ref{fig:20p} we show the ratio $w_{\rm p,early}/w_{\rm p,late} = b_{\rm early}/b_{\rm late}$ of the control samples.  Each pair consists of 138 early-analog and 121 late-analog clusters, identical to that of the $z_{\rm 20ex,Mz}$ set.
In the upper left part of each panel, we show the masses of the early-analogous and late-analogous clusters as measured via stacked WL (in unit of $10^{14} h^{-1}\,M_\odot$), and the probability for the pair to be drawn from the same parent population (using the clustering measurements within the mid-range).  
The red horizontal line denotes the bias ratio expected (in the absence of AB) given the masses of the cluster samples in consideration.  As we arrange the pairs in increasing masses, the red line is always very close to unity (shown as the black  dotted line).
None of these shows probabilities as low as that of the  $z_{\rm 20ex,Mz}$ set.

For these 14 pairs of cluster samples, we have fit an NFW profile to the spatial distribution of member galaxies.  For each pair, which by design have similar masses, we  compute the ratio of the measured concentration parameters (early-analog over late-analog), as well as that of the total number of member galaxies.  The mean ratio and standard deviation of concentration are found to be (1.19, 0.24), while those of the total galaxy number are (0.90, 0.13); the scatters of these quantities are comparable to those from the literature
(about $15\%$ for concentration and $30\%$ for galaxy number, e.g., \citealt{lin04,hennig17}), thus we can regard both quantities to be  consistent with unity, indicating that these pairs of cluster samples have galaxy properties largely consistent with each other.  Furthermore, we note that the mean separations between the BCG and  cluster center are $(0.20\pm 0.01)r_{200m}$ and $(0.22\pm 0.01)r_{200m}$ for the early- and late-analog samples, respectively.

Finally,  the median magnitude gaps are found to be $\Delta_{12}=0.42\pm 0.01$ ($\Delta_{14}=0.99\pm 0.01$) for the 14 early-forming analog cluster samples (with a scatter of 0.05), while those for the late-forming analog ones are $\Delta_{12}=0.42\pm 0.01$ ($\Delta_{14}=0.97\pm 0.01$), also with a scatter of 0.05. 
These smaller magnitude gaps, as compared to those of our $z_{\rm z20ex,Mz}$ set (c.f.~ the values reported at the penultimate paragraph of Section~\ref{sec:gal}), provide further support that the formation time derived from {\it Elucid} is informative.

\section{Discussion} 
\label{sec:foundation}

\begin{table*}
	\centering
	\caption{Expected Observational Signatures of Various Possibilities of AB Detection}
	\label{tab:logic}
	\begin{tabular}{r|cccc} 
		\hline
		Observable &True AB & Observed  & Spurious AB due to & Spurious AB due to\\
		                   &              &   Trend       & Circularity  & Incorrect Cluster Mass \\
		\hline
Concentration & $c_{\rm e} > c_{\rm l}$ & $c_{\rm e} > c_{\rm l}$ & $c_{\rm e} = c_{\rm l}$ & $c_{\rm e} \approx  c_{\rm l}$\\
Galaxy Number & $N_{\rm e} < N_{\rm l}$ & $N_{\rm e} < N_{\rm l}$ & $N_{\rm e} =N_{\rm l}$ & $N_{\rm e} \lesssim N_{\rm l}$\\
BCG Offset & $d_{\rm e} < d_{\rm l}$ & $d_{\rm e} < d_{\rm l}$ & $d_{\rm e} = d_{\rm l}$ & $d_{\rm e} > d_{\rm l}$\\
Magnitude Gap & $\Delta_{\rm e} > \Delta_{\rm l}$ & $\Delta_{\rm e} > \Delta_{\rm l}$ & $\Delta_{\rm e} = \Delta_{\rm l}$ & $\Delta_{\rm e} > \Delta_{\rm l}$\\
		\hline
	\end{tabular}
\end{table*}

As our $z_{\rm 20ex,Mz}$ set  is constructed using
halo formation time from  {\it Elucid}, some readers may wonder whether we are measuring the AB signal in {\it Elucid} or in the real Universe?  A related concern is, would it be circular to use a CDM-based simulation (which must contain the AB signal) to infer the halo formation time, and in turn use it to 
construct the cluster samples?

In simplest terms, we can recast our study as a hypothesis test, with the ultimate goal of ruling out the null hypothesis that the real Universe has no AB.
Thinking of this problem in a Bayesian way, we have
\begin{equation}
P({\rm AB}\, |\, {\rm data}, Elucid) \propto P({\rm data}\, |\, {\rm AB}, Elucid) P({\rm AB} | Elucid)
\label{eq:bayes}
\end{equation}
where ``AB'' stands for ``AB exists in the Universe'', ``data'' refers to properties of our $z_{\rm 20ex,Mz}$ set (including WL mass, clustering measurements, as well as cluster galaxy properties), and the prior $P({\rm AB} | Elucid)$ should be taken to be uninformative, say 50\% chance for $P({\rm AB} | Elucid)=1$ or 0.  As for the likelihood $P({\rm data} | {\rm AB}, Elucid)$,
logically, we can consider four general cases that result from the combination of (1) whether there is AB in the real Universe (which is assumed to be described by the $\Lambda$CDM model\footnote{Given that AB is an important feature of $\Lambda$CDM, this phenomenon should naturally exist in the Universe.  One may then ask what is the point of carrying out the analysis presented in this paper?  Firstly, 
having a measurement that is consistent with a theoretical prediction is an important step in scientific analyses.  A prime example is the black hole shadow images obtained by the Even Horizon Telescope collaboration \citep{eht19,eht22}, which spectacularly confirm the predictions of General Relativity. 
Secondly, to cite one of the most famous experiments in cosmology,
to detect the baryon acoustic oscillation (BAO) signal, one needs to first {\it assume} the CDM framework is correct, then  proceed to convert the BAO scale from an angular to a physical one (e.g., \citealt{eisenstein05}).}), and (2) whether there is AB in {\it Elucid}, and compare our observational findings (presented in Sections~\ref{sec:gal} \& \ref{sec:robust}) with the predicted observational results to deduce in which of the four cases we are in.  

First of all, we do see a strong signal of AB in {\it Elucid} (i.e., the open triangles in the lower right panel of Fig.~\ref{fig:z20mz}), so we can rule out two of the four possible cases.  Then the main issue is to determine whether 
our data also requires AB to be present in the real Universe.
To answer this question, let us refer to Table~\ref{tab:logic}, where in the first column we list four observables we have presented in Section~\ref{sec:gal}, and in subsequent columns we list in turn the expected behavior when there is AB in the Universe, our observed trends, 
the behavior when we are seeing an AB-like signal  
simply because of any potential circularity in our methodology, and the behavior when 
such a
signal is simply due to an incorrect cluster mass estimation (that is, for example, the mass of our late-forming cluster sample is underestimated by at least $2\sigma$, which has a 2.3\% chance to occur; please see Section~\ref{sec:z20ex}).

In a Universe where there is AB, we expect that the early-forming clusters to have a higher mass concentration than their late-forming counterparts, when their masses are the same.  Given that our WL measurements could not provide adequate constraints (Fig.~\ref{fig:corner}), we have to resort to the spatial distribution of cluster galaxies.  Assuming the {\it red} galaxy distribution follows that of the matter (e.g., \citealt{adhikari21}), we then expect to have $c_{\rm e} > c_{\rm l}$.
As for the  number of member galaxies $N$,  it is expected that $N$ in early-forming clusters should be smaller than that in the late-forming ones (see our argument at the end of the first paragraph in Section~\ref{sec:gal}), resulting in $N_{\rm e} < N_{\rm l}$.  Similarly, in older clusters, the BCGs are expected to be closer to the center  (hence $d_{\rm e} < d_{\rm l}$), and have a larger magnitude gap  (yielding $\Delta_{\rm e} > \Delta_{\rm l}$).  These expectations are all consistent with what we have observed with the  $z_{\rm 20ex,Mz}$ set, shown in the third column.

In the fourth column, we show the expectation for the case where we are simply measuring AB in {\it Elucid}, and that AB does {\it not} exist in the real Universe (let us for the moment allow for such an inconsistency with $\Lambda$CDM). 
By ``circularity'', we mean that, given the density field {\it Elucid} was tasked to reconstruct is based on the groups and clusters from the Y07 catalog, the {\it Elucid} halos are naturally expected to be located in large-scale environments (then indirectly, {\it bias}) similar to the real clusters used in our analysis. 
 In such a case, the formation time indicator we use ($z_{20}$) is only meaningful for the {\it Elucid} halos and has nothing to do with the real clusters;\footnote{In principle, $z_{20}$ from {\it Elucid} may still trace the formation history of the real clusters, but cluster properties in the real Universe would not be correlated with the formation time given that there is no AB in the Universe.}
for all four observables, they should therefore be similar/identical for the early- and late-forming clusters, since they have similar/identical masses.
The different behavior of the observables from the expectation shown in the second column
provides evidence against the notion that we are measuring AB in the simulation.
It is  important to note that our cross-correlation function measurements rely primarily on the spatial distribution of the SDSS main galaxy sample, which is far below the spatial resolution of Elucid, so our measurements are not purely based on {\it Elucid}.
Another point worth noting is that the 
halo mass estimates from Y07
may not be accurate even in the cluster regime.  This is not only reflected in the late-forming sample of the $z_{\rm 20ex,Mz}$ set, but also the wide range of halo masses of our control sample clusters: recall that in Section~\ref{sec:robust}, these samples are all chosen to have similar Y07 halo mass distribution  as the $z_{\rm 20ex,Mz}$ set.
{\it It is thus highly nontrivial that $z_{20}$ from  Elucid can be used to split the real clusters into early- and late-forming ones that exhibit an AB-like signature.} 
\footnote{Despite the observational evidence, one might still argue, regarding any potential circularity in our methodology, that if we were to randomly shuffle $z_{20}$ among the {\it Elucid} halos, while all the large-scale structures in the local Universe are still preserved, we would not be able to see the AB signal.  
We believe such an argument is flawed, as a modified {\it Elucid} with randomized halo formation history would violate the basic constraints of the CDM model of cosmology (or likely any sensible models of structure formation): the time continuity of the density field would be broken, and the spatial power spectrum and high-order statistics would be incorrect at any epochs other than $z=0$.}

Finally, in the fifth column, we consider the case where an AB-like signal arises due to a significant underestimation of the mass of our late-forming cluster sample.  Given the weak-to-no dependence on cluster mass of the observed total cluster galaxy concentration (e.g., \citealt{lin04,hennig17}), we thus expect $c_{\rm e} \approx c_{\rm l}$ for the cluster mass range we consider (i.e., $\approx 0.9$ to $2\times 10^{14} h^{-1}\,M_\odot$).  For the galaxy number, given the observed trend that $N\propto M_{200m}^{0.8}$ with a scatter of about 30\% \citep{lin04}, it is expected that $N_{\rm e} \lesssim N_{\rm l}$, although it could easily be $N_{\rm e} \gtrsim N_{\rm l}$ given the scatter in the $N$--$M$ relation. 
We caution that the concentration and galaxy number derived from photometric galaxy data could suffer from projection effects (caused by large-scale structures along the line-of-sight; e.g., \citealt{zu17,sunayama19}), and therefore one has to be careful in interpreting these results.
We note that our cluster samples are based on spectroscopic redshifts, so the identification of clusters is not affected by projection effects.

For the BCG offset, from our control sample that contains 28 cluster samples with WL mass measurements, we see a clear trend of a decreasing offset with increasing cluster mass, and thus $d_{\rm e} > d_{\rm l}$ is expected.  Finally,  the magnitude gap is found to decrease with cluster mass \citep[e.g.,][]{yang08,lin10}, so we have $\Delta_{\rm e} > \Delta_{\rm l}$.
Comparing the expectations shown in this column with those shown in the second column, we see that the main distinction comes from the BCG offset.  
Given that the mean values of the observed $d_{\rm e}$  and $d_{\rm l}$ of the $z_{\rm z20ex,Mz}$ set differs at $3\sigma$ level, and are far smaller than those of the control samples, we believe this feature is robust;
taking into account the observed behavior of the galaxy concentration and galaxy number, 
 it is unlikely that what we have observed is due to incorrect mass estimation.
It is worth noting that none of the 14 pairs of control samples ``passes'' these observational tests.
Finally, we recall that even if the mean cluster mass of our late-forming sample is severely underestimated, the difference in the bias is still far from sufficient to explain the large difference in the large-scale biases of the early- and late-forming samples (Section~\ref{sec:z20ex}), strongly hinting at the presence of some AB-like mechanisms at work.

In conclusion, although we construct the $z_{\rm 20ex,Mz}$ set using the formation history from a constrained simulation, not directly observed halo properties, seeing both the AB-like large-scale bias ratio and the consistent trends of cluster galaxy properties requires AB to not only exist in {\it Elucid} but also in the real Universe.

\section{Summary and Prospects} 
\label{sec:disc}

Using a novel approach of combining a constrained simulation of the local universe {\it Elucid} with a sample of galaxy clusters from SDSS, we have constructed a pair of early- and late-forming cluster samples.  The formation time indicator $z_{20}$ of the clusters is derived from their counterpart massive halos found in {\it Elucid}.  While WL-based cluster mass estimates indicate the two samples are comparable  (within $1\sigma$), their large-scale biases differ 
significantly,
 indicates the presence of
AB  
(with the correct sign that older clusters are {\it less} biased than the younger ones).  Furthermore, the properties of the galaxy populations of the two samples are also consistent with the expectation of the age difference: 
the early-forming clusters have a more concentrated red galaxy surface density profile, much smaller number of member galaxies, smaller offset of BCGs from cluster center, and a larger magnitude gap, compared to the late-forming ones.
The signal is found to be consistent with the theoretical prediction of $\Lambda$CDM, which is remarkable given that there is no guarantee that the formation time of halos from {\it Elucid} would match that of the real clusters.
In the  unlikely ($\sim 2.5\%$ chance) case where the differences in clustering measurements are largely due to the cluster mass difference, and take into account of possible uncertainties in the theoretical predictions of large-scale bias, our cluster samples still exhibit an AB-like signal at  $\approx 3\sigma$ level.

Our analysis can also be regarded 
as a hypothesis test -- based on the extensive discussion presented in Section~\ref{sec:foundation}, we can  rule out the null hypothesis that the real Universe has no AB at 99.7\% confidence level.

 Our $z_{\rm 20ex,Mz}$ set would be invaluable for further studies of aspects related to AB, such as the measurements of the splashback radius \citep[e.g.,][]{more16}, pressure profile of the intracluster medium, the BCG-to-total stellar mass ratio, etc.
 Furthermore, the member galaxy properties of our cluster samples can serve as foundations for a natural extension of our pursuit
 towards a more empirical detection of AB,
 namely to use cluster samples purely constructed from observables.  One can devise ways to split clusters of similar masses into old and young ones based on the combination of spatial distribution of member galaxies, member galaxy number, BCG offset, and magnitude gap. 
To minimize projection effects, it is desirable to use large cluster samples built by spectroscopic redshifts (e.g., Y07,  \citealt{tinker21b}).

Forward-modeling techniques like the one used to construct {\it Elucid}, as well as BORG  \citep{jasche13}, TARDIS \citep{horowitz19}, COSMIC BIRTH \citep{kitaura21}, are gaining increasing popularity (e.g., \citealt{nguyen20,tsaprazi21,ata22}), and they have immense potential for advancing our understanding of structure formation.
In this study we have demonstrated that a pair of cluster samples exhibiting a probable AB signal can be constructed using {\it Elucid}. 
With {\it Elucid}, in principle we can further study AB manifested in other halo properties such as spin or concentration. Furthermore, 
\citet{yang18} have also used {\it Elucid} to develop the ``neighborhood abundance matching'' method.
Finally, with rich spectroscopic data coming from surveys like Dark Energy Spectroscopic Instrument \citep[DESI;][]{desi16} and Subaru Prime Focus Spectrograph \citep[PFS;][]{takada14,greene22}, one can expect the reconstruction to become much more reliable, particularly at higher redshifts (e.g., \citealt{ata22}), which would facilitate studies of AB 
in the distant universe.

To enhance the legacy value of {\it Elucid,}
with this paper we  make  the {\it Elucid} data 
publicly available\footnote{\url{https://gax.sjtu.edu.cn/data/ELUCID.html}}.

\begin{acknowledgements}
We are grateful to Huiyuan Wang for providing data from the {\it Elucid} simulation, and Rachel Mandelbaum for sharing the SDSS shape catalog.
We thank an anonymous referee for comments that have improved the clarity of the paper.
We appreciate helpful comments from Ue-Li Pen, Zheng Zheng, Yao-Yuan Mao, Huiyuan Wang, Xiaohu Yang, Houjun Mo,  Eiichiro Komatsu, Yen-Chi Chen, Benedikt Diemer, Andrey Kravtsov, Neal Dalal,
and You-Hua Chu. 
YTL is supported by the National Science and Technology Council of Taiwan under grants MOST 111-2112-M-001-043, MOST 110-2112-M-001-004, and MOST 109-2112-M-001-005, and a Career Development Award from Academia Sinica (AS-CDA-106-M01). 
 HM is supported by World Premier International Research Center Initiative, MEXT, and JSPS KAKENHI Grant Numbers JP20H01932 and JP21H05456.
 HG is supported by NSFC (Nos.~11922305, 11833005).
 TWL is supported by the MOST 111-2112-M-002-015-MY3, the Ministry of Education, Taiwan (Yushan Young Scholar grant NTU-110VV007), National Taiwan University research grant (NTU-CC-111L894806).
YTL thanks IH, LYL and ALL for constant encouragement and inspiration.

Funding for the SDSS and SDSS-II has been provided by the Alfred P.~Sloan Foundation, the Participating Institutions, the National Science Foundation, the U.S.~Department of Energy, the National Aeronautics and Space Administration, the Japanese Monbukagakusho, the Max Planck Society, and the Higher Education Funding Council for England. 
The SDSS Web Site is http://www.sdss.org/.

The SDSS is managed by the Astrophysical Research Consortium for the Participating Institutions. The Participating Institutions are the American Museum of Natural History, Astrophysical Institute Potsdam, University of Basel, University of Cambridge, Case Western Reserve University, University of Chicago, Drexel University, Fermilab, the Institute for Advanced Study, the Japan Participation Group, Johns Hopkins University, the Joint Institute for Nuclear Astrophysics, the Kavli Institute for Particle Astrophysics and Cosmology, the Korean Scientist Group, the Chinese Academy of Sciences (LAMOST), Los Alamos National Laboratory, the Max-Planck-Institute for Astronomy (MPIA), the Max-Planck-Institute for Astrophysics (MPA), New Mexico State University, Ohio State University, University of Pittsburgh, University of Portsmouth, Princeton University, the United States Naval Observatory, and the University of Washington.

\end{acknowledgements}

\bibliographystyle{aa}

\end{document}